\documentclass{PoS}
\usepackage{url}

\title{Measurement of the energy flow in a large $\eta$ range and forward jets at LHC at $\sqrt{s} =$ 0.9 TeV, 2.36 TeV and 7 TeV}

\ShortTitle{Energy Flow in a large $\eta$}

\author{\speaker{Sercan Sen}\thanks{For CMS Collaboration.}\\
	University of Iowa, Iowa City, IA 52242-1479, U.S.A.\\
        E-mail: \email{Sercan.Sen@cern.ch}}

\abstract{A measurement is presented for the energy flow of minimum bias events in the forward region (3.15 $< \mid\eta\mid <$ 4.9,
where $\eta$ is the pseudorapidity) of the CMS detector at the LHC for center-of-mass
energies $\sqrt{s} =$ 0.9 TeV, 2.36 TeV and 7 TeV. The measurement is compared to Monte Carlo simulations, 
which use a model of multiparton interactions for the underlying event. 
In addition, production of forward jets was studied for the very first 
$pp$ collisions at $\sqrt{s} =$ 0.9 TeV at LHC.}

\FullConference{XVIII International Workshop on Deep-Inelastic Scattering and Related Subjects, DIS 2010\\
		April 19-23, 2010\\
		Firenze, Italy}

\begin{document}

\section{Introduction}

At very large center-of-mass energies, the momentum fraction of the proton carried by the partons in the hard scattering (x1, x2)
can become very small. In such a small-x region, parton densities might become very large and the probability 
of more than one partonic interaction per event increases as depicted in Fig 1. This approach is described in the
models of multiparton interactions (MPI)~\cite{MPI}, however, the details of the energy dependence of MPI cross sections is not well known yet. 

\begin{figure}[!ht]
  \begin{center}
    \includegraphics[width=0.35\textwidth]{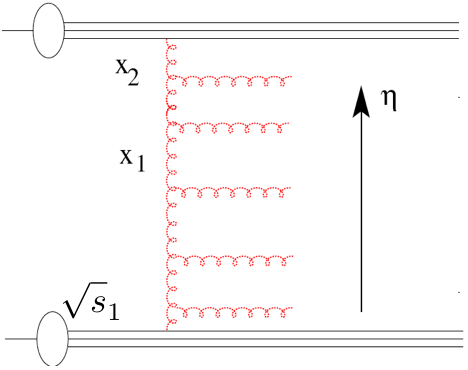}\hspace{1cm}\includegraphics[width=0.35\textwidth]{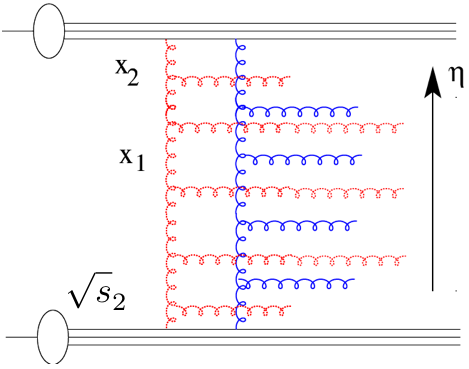}\\
(a)      \hspace{6cm}   (b)
%     \hspace{1cm}
    \caption{(a) Radiated partons coming from a parton shower. (b) More than one partonic interaction chain.}
\label{fig:1}
  \end{center}
\end{figure}

The measurement of the underlying event using the energy flow in the forward region is complementary to the
measurement of the underlying event in the central region, thus can provide
additional input to the determination of the parameters for the MPI models~\cite{MPI2}.
Forward jet production at the Large Hadron Collider (LHC) is also an ideal process to investigate small-x QCD effects \cite{smallx}. 
In this study, a first measurement of forward jets and the energy flow in the forward region 
(3.15 $< |\eta| <$ 4.9) of the Compact Muon Solenoid (CMS) detector is presented. 

\section{CMS Hadron Forward Calorimeter}

CMS \cite{CMS} is a general-purpose detector designed to run at the highest 
luminosity provided by the CERN LHC. 
Two hadron forward calorimeter (HF$\pm$), one on each side of the CMS interaction point (IP),
at about $\pm$11 m, cover the very forward angles of CMS, in the pseudorapidity range $3 < |\eta| < 5$.
Due to the severe radiation environment around beam pipe, HF calorimeters are built with radiation hard quartz fibers embedded in steel absorbers.
The signal in HF is produced in the form of \u{C}erenkov light generated by the showering particles passing through quartz fibers. 
Half of the fibers run over the full depth of the absorber whereas the other half, read out separately, start at a depth of 22 cm
from the front face of the detector. 
This structure makes it possible to distinguish showers generated by electrons and photons from those generated by hadrons.  
The objectives with the HF detector are to improve the measurement of the transverse energy and to measure very forward jets. 
The performance and thechnical details of the HF calorimeters can be found elsewhere~\cite{HF}. 

\section{Data Samples and Event Selection}

The data from $pp$ collisions, which were collected with the CMS detector at $\sqrt{s} =$ 0.9 TeV and 2.36 TeV in the fall of 2009 and 
at $\sqrt{s} =$ 7 TeV in March 2010 were used in this analysis. 

Two subsystems, the Beam Pick-up Timing for eXperiments (BPTX) and the Beam Scintillator Counters (BSC) \cite{CMS,BSC}, 
were used in the trigger of the detector readout.
The two BPTX devices, which are located around the beam pipe at a distance of $\pm$175 m from the IP, 
are designed to provide precise information on the structure and timing of the LHC beams, with a 
time resolution better than 0.2 ns. The two BSCs, consisting of a set of 16 scintillator tiles, 
are located along the beam line on each side of the IP at a distance of $\pm$10.86 m and they provide information 
on hits and coincidence signals with a time resolution of 3 ns.
The events analyzed in this study were selected by requiring the trigger signal in the BSC counters to be in coincidence with BPTX signals from both beams.
This requirement refers to the minimum bias trigger condition and indicates that there was activity in the forward regions, furthermore, 
at least one primary vertex was required to be reconstructed from at least 5 tracks with a $z$ distance 
to the interaction point with $|z|\leq$ 15 cm. These requirements ensure that the event is a collision candidate.

\section{Results}

\subsection{Forward Jets}

We searched for forward jets ($3 < |\eta| < 5$) in the very first $pp$ collisions data which were collected at $\sqrt{s} =$ 0.9 TeV at LHC~\cite{EventDisplay}. In Fig. 2,
an event is presented which has one forward jet and one backward jet with a corrected jet $p_{T} >$ 10 GeV. 

\begin{figure}[!Hhtb]
  \begin{center}
    \includegraphics[width=0.80\textwidth]{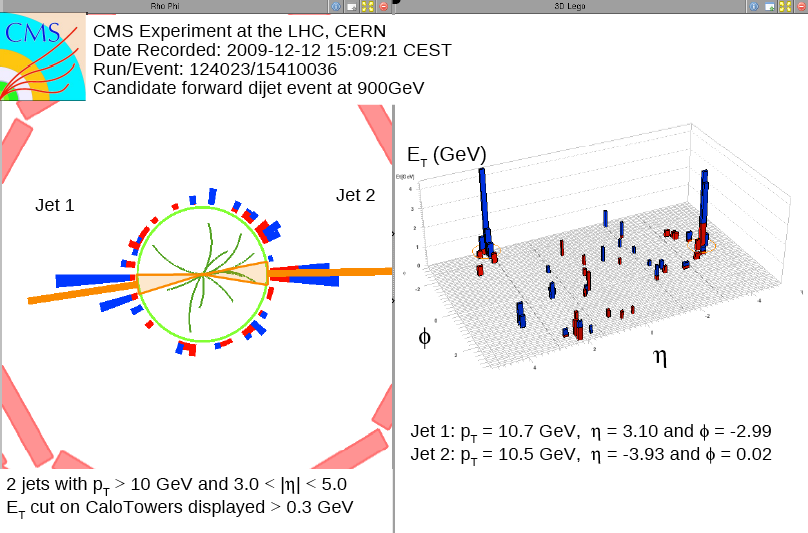}
%     \hspace{1cm}
    \caption{Display of an event with two forward jets. Data were collected by CMS in 2009 at $\sqrt{s} =$ 0.9 TeV $pp$ collisions.}
    \label{fig:2}
  \end{center}
\end{figure}

\subsection{Energy Flow}

The energy flow in the pseudorapidity region 3.15 < $|\eta|$ < 4.9 is measured for minimum bias events, and the ratio of energy flow is given by

\begin{equation}
R^{\sqrt{s_1},\sqrt{s_2}}_{Eflow} = \frac{\frac{1}{N_{\sqrt{s_1}}}\frac{dE_{\sqrt{s_1}}}{d\eta}}{\frac{1}{N_{\sqrt{s_2}}}\frac{dE_{\sqrt{s_2}}}{d\eta}}
\end{equation}
where $dE_{\sqrt{s}}$ is energy deposition in a region $d\eta$ (integrated over azimuthal angle $\phi$),
$N_{\sqrt{s}}$ corresponds to the number of selected minimum bias events, $\sqrt{s_1}$ refers to either 2.36 TeV or 7 TeV 
and $\sqrt{s_2}$ refers to 0.9 TeV.

\begin{figure}[!ht]
  \begin{center}
    \includegraphics[width=0.40\textwidth]{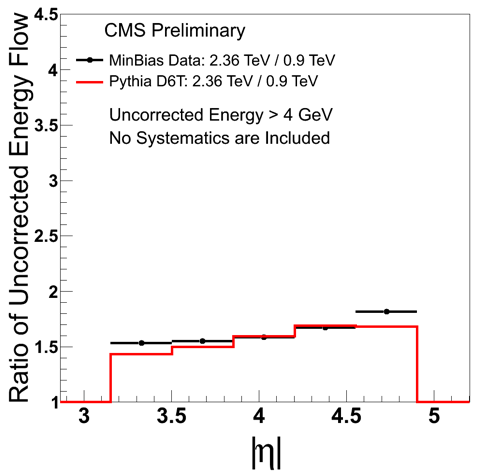}\hspace{1cm}\includegraphics[width=0.40\textwidth]{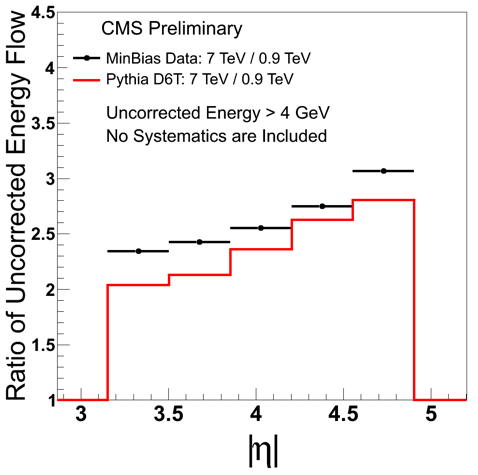}\\
	\hspace{6cm}
%     \hspace{1cm}
    \caption{The ratio of energy flow for $\sqrt{s_1} =$ 2.36 TeV to $\sqrt{s_2} =$ 0.9 TeV (left), and 
for $\sqrt{s_1} =$ 7 TeV to $\sqrt{s_2} =$ 0.9 TeV (right) as a function of $\eta$. Uncorrected data
are shown as points and the red line is the prediction from PYTHIA tune D6T~{\protect \cite{Pythia,PythiaManual}}.}
\label{fig:3}
  \end{center}
\end{figure}

In Fig. 3, the ratio of energy flows for different collision energies is shown as the average of the 
HF(+) and HF(-) responses~\cite{DetPer}. The pseudorapidity region 3.15 < $|\eta|$ < 4.9 is divided into 5 bins following 
the transverse segmentation of the HF calorimeters. The energy deposition in a given $d\eta$ is integrated over $\phi$ and 
4 GeV threshold is applied on the measured energy to be included in $dE/d\eta$ calculation. 
This threshold ensures that the noise is removed. Uncertainties due to systematic effects are not shown. Also there is no correction applied on energy scale.

\section{Conclusions}

In this study, an analysis of the very first $pp$ collisions data collected at LHC is presented. Forward jets, at $|\eta|>3$, 
have been observed for the first time 
in hadron-hadron collisions at this forward rapidities. Ratio of energy flow for minimum bias events at different $\sqrt{s}$ 
has been measured at the detector level. 
The results indicate that the energy flow is larger at forward rapidities at higher center-of-mass energies. 

At the time of writing, new measurements on minimum bias events and events having a hard scale
defined by a dijet with $E_{T,jet} >$ 8 GeV ($E_{T,jet} >$ 20 GeV for $\sqrt{s} =$ 7 TeV) in $|\eta| <$ 2.5 have been studied. 
Systematic effects are included and the results are compared to predictions from Monte Carlo event generators that were 
tuned to describe the charged particle spectra seen at central rapidities. A detailed description is under preparation. 

\section*{Acknowledgments}

I thank T. Yetkin for useful discussions throughout this work, G. Brona, H. Jung and K. Piotrzkowski for useful feedback and suggestions, Y. Onel for his support.

\end{document}